# Recent results in Scalar QED[*]

M. Baig[a] and J.B. Kogut[b]

[a]Grup de Física Teòrica, IFAE
Universitat Autònoma de Barcelona
08193 Bellaterra (Barcelona) SPAIN

[b]Physics Department
1110 West Green Street
University of Illinois
Urbana, Il. 61801/380 USA

In this talk we present some recent and new results on the phase structure and the continuum limit of lattice formulation of pure gauge QED coupled to Higgs fields. We show the presence of second order phase transition lines allowing for a continuum limit. Nevertheless, finite size analysis shows that this theory is logarithmically trivial, in the same sense than $\lambda \phi^4$ The role of magnetic monopoles is also shown.

## 1. INTRODUCTION

The results presented here come from a collaboration started in 1988. His first results have been published in [1–3]. The initial purpose of the collaboration was to establish the phase diagram of the scalar QED with non compact gauge fields coupled to Higgs fields. Once shown than this phase diagram contains second order phase transition lines, the main purpose has been to study the critical exponents of the transition in order to identify the continuum limit of the theory.

## 2. THE PHASE DIAGRAM

### 2.1. The action

Our model has gauge fields in the links of the lattice that are elements of the $U(1)$ gauge group $\theta_{x,\mu}$ and matter fields on the sites $\phi_x = e^{i\alpha(x)}$. The action is

$$S = \frac{1}{2}\beta \sum_p \theta_p^2 - \lambda \sum_{x,\mu}(\phi_x^* U_{x,\mu}\phi_{x+\mu} + cc), \qquad (1)$$

where $\theta_p$ is the circulation of the gauge field around a plaquette, $\beta = \frac{1}{e}$ is the gauge coupling and $\lambda$ the Higgs coupling.

We have chosen this action because:

[*]Presented by M. Baig

- Phenomenologically one wants to understand the Higgs mechanism in the standard model. Scalar QED seems to be a good place to do that.

- Compact scalar QED seems to have only first order transition lines, and it does not allow for a continuum limit (except in an end-point).

- The use of higgs fields with unitary norm simplifies the computation and it is believed to lie in the same universality class than the ordinary Higgs model[4].

### 2.2. Measurements

We have measured both, the internal energies of the gauge and Higgs parts of the action,

$$E_\gamma = \frac{1}{2} < \sum_p \theta_p >, \qquad (2)$$

$$E_h = < \sum_{x,\mu} \theta_x U_{x,\mu}\theta_{x+\mu} + cc >, \qquad (3)$$

and also the corresponding specific heats.

The phase structure is represented in Fig.1. Basically, one finds only two phases, Coulomb and Higgs, separated by a second order transition line, that becomes of first order in the $\gamma \to \infty$ limit.

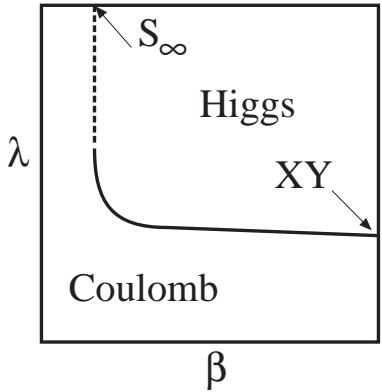

Figure 1. Qualitative phase diagram of the non-compact scalar QED.

## 3. THE CONTINUUM LIMIT

In order to establish the continuum limit of this theory a finite size analysis must be done over the phase transition. The "standard" expectations for the maximum of the specific heat and the location of coupling are

$$C_{max}(L) \sim L^{\frac{\alpha}{\nu}}, \qquad (4)$$

$$g_c(L) - g_c \sim L^{-\frac{1}{\nu}}. \qquad (5)$$

However, the numerical evidence does not support this picture, even for mean field critical exponents. Moreover, there are other possibilities for the critical behavior, as that suggested by the $\lambda\phi^4$ theory, witch predicts

$$C_{max}(L) \sim (\ln L)^p, \qquad (6)$$

$$g_c(L) - g_c \sim \frac{1}{L^2(\ln L)^q}. \qquad (7)$$

We have done numerical simulations over a definite line of the phase diagram that crosses the phase transition. For instance, we have fixed the gauge coupling $\beta = 0.2$. We have studied lattices from $6^4$ to $20^4$ using statistics of the order of 20 millions of sweeps for each size. We have measured both specific heats

$$C_\gamma = \frac{1}{4L^4}(<E_\gamma^2> - <E_\gamma>^2), \qquad (8)$$

$$C_h = \frac{1}{4L^4}(<E_h^2> - <E_h>^2), \qquad (9)$$

and also the usual Binder cumulant for each energy.

The results show the confirmation of the logarithmic behavior of the critical exponents giving, for instance, a scaling law for the gauge specific heat

$$C_\gamma^{max}(L) = a(\ln L)^\rho + b, \qquad (10)$$

giving $\rho = 1.4(2)$ with a 90% of confidence level. The quality of the data can be seen in Fig 2.

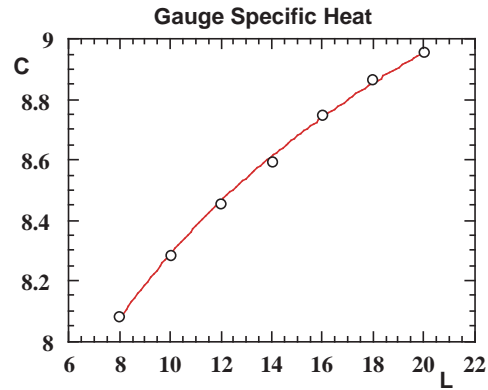

Figure 2. Maximum of the gauge specific heat vs. lattice size.

All these results support the conclusion that the scalar QED with non compact gauge fields is a logarithmically trivial theory.

## 4. THE DYNAMICS

In 1977, Banks, Myerson and Kogut[5] showed that QED can be considered as a theory of photons and monopoles. In 1980, numerical simulations[6,7] showed a monopole-condensation phenomena just over the phase transition of the pure gauge compact QED. More recently, it has been found a monopole percolation phenomena in both, the full non-compact QED with fermions, and the pure gauge non compact case[8,9].

## 4.1. The non-compact case

Following these lines, we have measured the behaviour of the magnetic monopoles for the case of non-compact scalar QED. The results show the presence of a line of critical behavior on the plane $(\beta, \lambda)$, line that is connected with the monopole-percolation observed in the pure gauge non-compact study. In Fig. 3 it is represented the critical behavior of the monopole susceptibility for a measurement over the line $\lambda = 0.23$. The critical exponent is compatible with that of the pure gauge case. This shows that for the case of scalar matter the monopole-percolation transition is decoupled of the energy phase transition line.

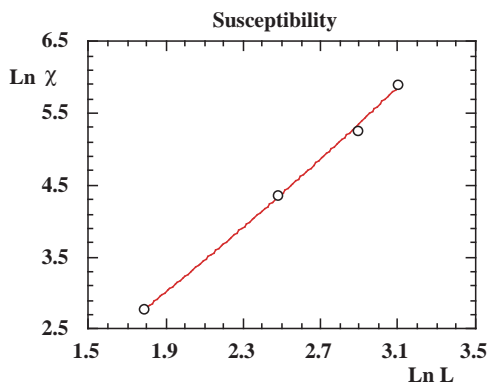

Figure 3. Maximum of the monopole susceptibility vs. lattice size.

## 4.2. The compact case

It would be very instructive to compare the behavior of monopoles in both, compact and non-compact cases. To this end, we have performed simulations on the pure gauge QED with the compact action looking for the behaviour of the monopole-percolation parameters, i.e. the monopole susceptibility and the $M = \frac{n_{max}}{n_{tot}}$ parameter. The results obtained up to now show a clear discontinuity of both parameters just over the first order transition point in such a way that they can be considered as a good order parameters to study this transition[10].

## 5. CONCLUSIONS

We have analyzed the lattice formulation of the scalar Quantum Electrodynamics using a non-compact gauge fields and unitary Higgs fields. The phase diagram shows the presence of a line of second order phase transitions. An accurate finite size analysis on larger lattices (from L=6 to 20) using a reasonable statistics shows a scaling behaviour compatible with logarithmic triviality, in the same sense than $\lambda \phi^4$. The analysis of the behavior of magnetic monopoles shows a percolation-phenomena, like the non compact pure gauge case, being decoupled from the phase transition line.

The simulations done here used mainly the CRAY C90 at PSC and NERSC. Part of the simulations have been done also on the CRAY-YMP of CESCA (Centre de Supercomputació de Catalunya). One of us (MB) acknowledges the support from CESCA and CICYT.